\documentstyle [12pt]{article}

\begin{document}
\def \inbar{\vrule height1.5ex width.4pt depth0pt}
\def \C{\relax\hbox{\kern.25em$\inbar\kern-.3em{\rm C}$}}
\def \R{\relax{\rm I\kern-.18em R}}
\newcommand{\Z}{\ Z \hspace{-.08in}Z}
\newcommand{\be}{\begin{equation}}
\newcommand{\ee}{\end{equation}}
\newcommand{\bea}{\begin{eqnarray}}
\newcommand{\eea}{\end{eqnarray}}
\newcommand{\nn}{\nonumber}
\newcommand{\kt}{\rangle}
\newcommand{\br}{\langle}
\newcommand{\lll}{\left( }
\newcommand{\rrr}{\right)}

\baselineskip=24pt

\title {Lattice Topological Field Theory on Non-Orientable Surfaces}
\vspace{1cm}
\author {V. Karimipour and  A. Mostafazadeh\\ \\
Institute for Studies in Theoretical Physics and Mathematics\\
P.~O.~Box 19395-5746, Tehran, Iran \\ Department of Physics , Sharif
University of Technology \\P.~O.~Box 11365-9161, Tehran, Iran}
\date{}
\maketitle

\def\en{end{equation}}

\vspace{2cm}

\begin{abstract}
The lattice definition of the two-dimensional topological quantum
field theory [Fukuma, {\em et al}, Commun.~Math.~Phys.\ {\bf 161},
157 (1994)] is generalized to arbitrary (not necessarily orientable)
compact surfaces. It is shown that there is a one-to-one correspondence
between real associative $*$-algebras and the topological state sum invariants
defined on such surfaces. The partition and $n$-point functions on
all two-dimensional surfaces (connected sums of the Klein bottle or
projective plane and $g$-tori) are defined and computed for arbitrary
$*$-algebras in general, and for the the group ring $A=\R[G]$
of discrete groups $G$, in particular.
\end{abstract}

\newpage

\section{Introduction}
The introduction of topological field theory (TFT) by Witten
[W1,W2], its axiomatization by Atiyah [A], and the novel approach
of employing the TFT techniques to attack problems of topology and geometry
[W1,W2,MS,DW], have motivated many authors to provide tools for
rigorous construction of TFT models [TV,D,DJN,KS].

In the framework of lattice topological field theory (LTFT),
a rigorous construction should inevitably start with a triangulation
of the manifold under consideration. In three dimensions (resp.~two
dimensions) the basic observation [TV] (resp.~[FHK,BP])
has been that the $6j$-symbols of $U_q(sl(2,\C))$ and a large class
of other algebras (resp.~ structure constants of associative algebras)
obey the symmetries of a tetrahedron (resp.~ triangle) and satisfy
identities which may be interpreted geometrically in terms of glued
tetrahedra (resp.~ triangles). Associating state sums (partition
functions) with a triangulation, one could show that the partition
function is independent of the triangulation, i.e., it is a topological
invariant.

In the basic definition of TFT [A], which is motivated by the path
integral examples of Witten, and in the lattice models constructed
afterwards, the orientability of the underlying manifold plays a
crucial role. To the best of our knowledge, state sum models on
non-orientable manifolds have not yet been constructed, even in two
dimensions. The aim of the present paper is to construct, in general terms,
topological state sums (partition functions) and observables on non-orientable
two-dimensional surfaces.

In our opinion, this direction of generalization of TFT deserves
consideration for two reasons. The first of these is a possible relevace
of topological correlation functions on non-orientable surfaces to
the open string theory [GSW]. The second and a more fundamental
reason is that mathematically, topological invariants are well defined
for orientable as well as non-orientable manifolds, whereas the axioms
of TFT [A], which are based on the path integral formulation
of QFT, and the state sum models mentioned above, rely heavily on the
orientability of the manifold. Therefore, it is desirable to see if one can
generalize state sum invarinats to also cover the non-orientable
cases. Although our considerations are restricted to two dimensions,
our basic idea seems to be generalizable to three dimensions as well.

The paper is organized as follows: In Sec.~2, a brief review of LTFT
on orientable surfaces is presented. In Sec.~3, the definition of
state sums on non-orientable surfaces is given and the generalized
local (Matveev) moves are introduced accordingly. It is shown that
the state sums, so defined, are invariant under these moves, provided
that a set of consistency conditions are fulfilled. Thus, the state
sums are sensitive only to the topological properties of the surface.
In Sec.~4, it is shown how real associative $*$-algebras provide the
general solution of the consistency conditions.
In Sec.~5, the observables are defined and for all $*$-algebras
the correlations on all two-dimensional surfaces are calculated.
Sec.~6 is devoted to the study of a particular example where
the $*$-algebra is taken to be the group ring of a discrete group.

\section{Definition of LTFT on Oriented Surfaces [FHK]}

Let ${\bf \Sigma}$ be a closed oriented surface of genus $g$ , ${\bf T_g}$ a
triangulation of ${\bf\Sigma}$. Then the partition function of the lattice
model associated with ${\bf T_g}$ is defined as follows: First, for an
oriented triangle ${\bf\Delta}$ in ${\bf T_g}$, one makes a coloring
according to its orientation. That is, one gives a set of color
indices running from $1$ through $N$, to three edges of the triangle.
        \begin{figure}
        \vspace{1.5in}
        \caption{Colored triangle with complex valued $C_{abc}$'s.}
        \label{h2.1}
        \end{figure}
                \begin{figure}
                \vspace{1.5in}
                \caption{Gluing two triangles}
                \label{h2.2}
                \end{figure}
One then assigns a complex number $C_{abc}$ to a triangle with ordered
color indices $a, b, c $ (Fig.~\ref{h2.1}). Here it is assumed that
$C_{abc}$ is symmetric under cyclic permutations of the indices:
        $$ C_{abc} = C_{bca} = C_{cab}\;. $$
Note, however, that $C_{abc}$ is not necessarily totally symmetric.
Next, all the triangles of ${\bf T_g}$ are glued  by contracting their
indices with a metric $g^{ab}$ (Fig.~\ref{h2.2}). Thus one obtains a
complex valued function of $g^{ab}$ and $C_{abc}$ for each triangulation
${\bf T_{g}}$, and one interprets it as the partition function of the
lattice model, ${\cal Z}={\cal Z}({\bf T_g})$.

Alternatively the construction of the partition function can be done in the
dual graph ${\bf T_g^*}$ of $ {\bf T_g }$. Here one assigns $ C_{abc} $ to
the vertices and $ g^{ab} $ to the links (Fig.~\ref{h2.3}).
        \begin{figure}
        \vspace{1.5in}
        \caption{The propagator $g^{ab}$ and the three-point vertex
        in the dual diagram.}
        \label{h2.3}
        \end{figure}
One further assumes that $( g^{ab}) $ has its inverse $( g_{ab} )$; and raises
or lowers indices using these matrices. One should then choose the coefficients
$C_{abc}$ and $g^{ab} $ such that the partition function is invariant under
any local changes in the triangulation ${\bf T_g}$ or in the dual diagram
${\bf T_g^* }$.

A possible set of local moves which relates any two triangulations,
is the two-dimensional version of the Matveev moves. These are
the fusion transformation (Fig.~\ref{H2.4}) and the bubble transformation
(Fig.~\ref{H2.5}).
        \begin{figure}
        \vspace{2in}
        \caption{Fusion transformation in ${\bf T_g}^*$.}
        \label{H2.4}
        \end{figure}
                \begin{figure}
                \vspace{1in}
                \caption{Bubble transformation in ${\bf T_g}^*$.}
                \label{H2.5}
                \end{figure}
Fig.~\ref{H2.6} demostrates an interpretation of the fusion transformation
in the triangulation ${\bf T_g}$.
        \begin{figure}
        \vspace{1.5in}
        \caption{Fusion transformation in ${\bf T_g}$}
        \label{H2.6}
        \end{figure}
In Ref.~[FHK], it is claimed that the bubble transformation can be expressed
only in the dual diagram $ {\bf T_g^*}$. However, we would like to emphasize
that it also has a clear interpretation in ${\bf T_g}$.
The meaning of the bubble transformation becomes clear only when one combines
it with the fusion transformation. In fact, we can add a vertex to the left
side of both diagrams in Fig.~\ref{H2.5} and obtain Fig.~\ref{H2.7}.
Now we perform a fusion transformation in the right hand figure
to obtain Fig.~\ref{H2.8}. But this last equality is nothing but the
barycentric subdivision in ${\bf T_g}$ (Fig.~\ref{H2.9}).
        \begin{figure}
        \vspace{2in}
        \caption{Bubble transformation applied to a vertex in ${\bf T_g}^*$.}
        \label{H2.7}
        \end{figure}
                \begin{figure}
                \vspace{2in}
                \caption{Barycentric subdivision in ${\bf T_g}^*$}
                \label{H2.8}
                \end{figure}
                        \begin{figure}
                        \vspace{2in}
                        \caption{Barycentric subdivision in ${\bf T_g}$}
                        \label{H2.9}
                        \end{figure}

Invariance of the partition function ${\cal Z}({\bf T_g})$ under the first
and the second Matveev moves enforces the following constraints on the
parameters $C_{abc}$ and $g^{ab}$ respectively.
        \begin{equation}
        C_{ab}^{~~p} C_{pc}^{~~d} = C_{bc}^{~~p}C_{ap}^{~~d}
        \label{e2.1}
        \end{equation}
                \begin{equation} g_{ab} = C_{ac}^{~~d}C_{bd}^{~~c}  \;.
                \label{e2.2}
                \end{equation}

In Ref.~[FHK], it is shown that every semisimple associative
algebra $A$ provides a solution of these constraints.
The coefficients  $ C_{ab}^{~~c}$ are identified with the structure constants
of the associative algebra. In view of the definition of the
structure constants in terms of a basis $\{\phi_a~:~a=1,\cdots,N\}$:
        \begin{equation}
        \phi_{a} \phi_{b} = C_{ab}^{~~c} \phi_c \;,
        \label{e2.3}
        \end{equation}
Eq.~(\ref{e2.1}) is the expression of the associativity of the algebra,
whereas Eq.~(\ref{e2.2}) yields the metric $g_{ab}$ in terms of the
structure constants. Note that if we define $g_{ab}:=\br\phi_a,\phi_b\kt$,
then the cyclic symmetry of $C_{abc}$ is expressed by
        $$ \br \phi_a , \phi_b \phi_c  \kt = \br \phi_a \phi_b , \phi_c\kt $$
In order for $g_{ab}$ to have an inverse, the algebra should be semisimple.
One then has the following theorem [FHK]:
        \begin{itemize}
        \item[] {\bf Theorem~1}: {\em There is a one-to-one correspondence
        between the set of all LTFT's on orientable surfaces, as defined
        above, and the set of all semisimple associative algebras.}
        \end{itemize}
Note that if one considers the regular representation of the algebra A :
        \begin{equation}
        [\pi(\phi_a)]^c_{~b} = C_{ab}^{~~c} \;,
        \label{e2.4}
        \end{equation}
then one finds:
        \begin{equation}
        g_{ab} = tr ( [\pi(\phi_a)][\pi(\phi_b)] ) \;,
        \label{e2.5}
        \end{equation}
                \begin{equation} C_{abc} =
                tr ( [\pi(\phi_a)][\pi(\phi_b)][\pi(\phi_c)])\;.
                \label{e2.6}
                \end{equation}
The latter equations manifestly demonstrate the symmetry of $ g_{ab} $ and
the cyclic symmetry of $ C_{abc} $.

\section {Generalization of LTFT to Arbitrary Compact Surfaces}
Consider a closed (possibly non-orientable) surface ${\bf \Sigma}$, a fixed
triangulation ${\bf \Sigma}^\alpha$ of ${\bf \Sigma}$, and equip each triangle
of ${\bf \Sigma} ^\alpha$ with an
arbitrary orientation.\footnote{Note that here we use ${\bf \Sigma^\alpha}$
rather than ${\bf T_g}$ to denote a particular triangulation (indexed by
$\alpha$), for convenience.} Denoting the number of triangles of ${\bf
\Sigma}^\alpha$ by $F$, one has $2^F$ possible ways of assigning orientations
to the triangles. We shall call ${\bf \Sigma}^\alpha$
together with such an assignment a {\em locally oriented triangulation}
of ${\bf \Sigma}$. Locally oriented triangulations corresponding to
${\bf \Sigma^\alpha }$ are labeled by ${\bf\Sigma}^{\alpha,k},~~
k=1,2,\cdots,2^F$. We shall denote the set of all ${\bf \Sigma}^{\alpha,k}$'s
by
${\bf  \tilde\Sigma}^\alpha$ and the set of all locally oriented
triangulations of ${\bf \Sigma}$ by ${\bf \tilde\Sigma}$, i.e.,
        \bea
        {\bf  \tilde\Sigma}^\alpha&:=& \{ {\bf
        \Sigma}^{\alpha,k}~:~k=1,\cdots,2^F\}\;,         \nn\\
        {\bf  \tilde\Sigma} &:=&\cup_\alpha {\bf
        \tilde\Sigma}^\alpha\;.         \nn
        \eea

We shall construct the partition function as a real valued map
${\bf\cal Z}:\{ {\bf\Sigma^{\alpha k}}:\forall {\bf\Sigma},~\alpha,~k\}
\to\R$. By its topological invariance we mean that for a fixed surface
${\bf\Sigma}$, this map has a constant value on ${\bf \tilde \Sigma}$.
Topologically, this means that ${\cal Z}$ should be invariant under the
following local moves in  the space ${\bf  \tilde \Sigma } $:
        \begin{itemize}
        \item[A.]
        {\bf Flipping}:  With a fixed triangulation we can change
        the orientation of any arbitrary triangle and thereby move in the
        subsets  $ {\bf\tilde \Sigma}^\alpha $ .
        \item[B.]
        {\bf Matveev Moves}: These enable us to interpolate between different
        subsets $ {\bf  \tilde \Sigma}^{\alpha} $  and $ {\bf  \tilde
        \Sigma}^{\beta}$ in ${\bf\tilde\Sigma}$.
        \end{itemize}

To construct the partition function, we proceed as follows: To each
locally oriented triangle, carrying the color indices $a,b$, and $c$,
we assign a real number $C_{abc}$ according to the orientation of the
triangle (Fig~\ref{f2.1}).
        \begin{figure}
        \vspace{1.5in}
        \caption{Colored oriented triangle with real $C_{abc}$.}
        \label{f2.1}
        \end{figure}
Each pair of triangles with adjacent edges labeled by $a$ and $b$, are
glued together by means of contracting their indices using two types of
matrices: $g^{ab}$ or $\sigma^{ab}$, according to whether the orientations of
the adjacent triangles are compatible or not, respectively, (Fig.~\ref{f2.2}).
        \begin{figure}
        \vspace{1.5in}
        \caption{Gluing two oriented trangles.}
        \label{f2.2}
        \end{figure}
For brevity, we shall call two adjacent triangles with (in)compatible
orientations, {\em (in)compatible triangles}.

Consistency of this prescription requires $C_{abc}$ to be cyclically
symmetric, and $g^{ab}$ and $\sigma^{ab}$ to be symmetric in their indices:
        \bea
        C_{abc}&=&C_{bca}\:=\:C_{cab}\;,
        \label{e3.1}\\
        g^{ab}&=&g^{ba}\;,~~~~\sigma^{ab}\:=\:\sigma^{ba}\;.
        \label{e3.2}
        \eea

In the dual diagram, we associate a vertex to each triangle,
a double line (propagator) to each common edge of two compatible triangles
and a twisted double line (twisted propagator) to each common edge of two
incompatible triangles. Thus, the numbers $C_{abc}$, $g^{ab}$, and $
\sigma^{ab}$ are assigned to the vertices, propagators, and twisted
propagators, respectively, (Fig.~\ref{f2.3}).
        \begin{figure}
        \vspace{3in}
        \caption{Oriented vertices, twisted and untwisted propagators.}
        \label{f2.3}
        \end{figure}

Contracting all the indices, one obtains a real number which we interpret
as the partition function of the lattice model based on the locally oriented
triangulation ${\bf \Sigma^{\alpha,k}}$. The next step is to find out the
conditions on $C_{abc}\,,~g^{ab}$, and $\sigma^{ab}$ that would imply the
invariance of ${\cal Z}$ under flipping, i.e., ${\cal Z}={\cal Z}({\bf
\tilde{\Sigma}}^{\alpha})$, and Matveev moves, i.e., ${ \cal Z}={\cal Z}(
{\bf \tilde {\Sigma }})$.

Consider a locally oriented triangulation ${\bf \Sigma^{\alpha ,k}}$
and change the orientation of an arbitrary triangle in ${\bf \Sigma^{
\alpha ,k}}$ while the orientations of the rest of the triangles are kept
unchanged. In this case, one of the cases depicted in Fig.~\ref{f2.4} may
happen.
        \begin{figure}
        \vspace{3in}
        \caption{Flipping transformation in ${\bf T_g}$.}
        \label{f2.4}
        \end{figure}
In view of Fig.~(\ref{f2.4}), invariance of ${\bf {\cal Z}}$ under
flipping leads to the following relations:
        \bea
        g^{aa'}g^{bb'}g^{cc'}C_{a'b'c'}&=&
        \sigma^{aa'}\sigma^{bb'}\sigma^{cc'}C_{a'c'b'}\;,
        \label{e3.5}\\
        \sigma^{aa'}g^{bb'}g^{cc'}C_{a'b'c'}&=&
        g^{aa'}\sigma^{bb'}\sigma^{cc'}C_{a'c'b'}\;.
        \label{e3.6}
        \eea

Next, we require invariance of ${\bf \cal Z}$ under local Matveev Moves.
Consider an arbitrary pair of adjacent triangles. Without loss of
generality, we assign compatible orientation to this pair and perform
the first Matveev move (Fig.~\ref{f2.5}).
        \begin{figure}
        \vspace{1.5in}
        \caption{Fusion transformation in ${\bf T_g}$.}
        \label{f2.5}
        \end{figure}
Invariance of ${\bf \cal Z}$ under this move yields the following relation
for $C_{abc}$'s:
        \be
        C_{da}^{~~p} C_{pb}^{~~c}=C_{ab}^{~~p} C_{dp}^{~~c}\;.
        \label{e2.7}
        \ee
Next perform a barycentric subdivision of an arbitrary oriented triangle,
Fig.~\ref{f2.6}.
        \begin{figure}
        \vspace{1.5in}
        \caption{Barycentric subdivision in ${\bf T_g}$.}
        \label{f2.6}
        \end{figure}
This yields the following relation:
        \be
        g_{ab}=C_{ac}^{~~d}C_{bd}^{~~c}\;.
        \label{e2.8}
        \ee
Note that once we have chosen the orientation of the triangles, the
orientation of the new triangles obtained after affecting the
Matveev moves is not arbitrary. It is dictated by the external
edges of the subdiagram where the Matveev moves take place.

Thus we have shown that the conditions  Eqs.~(\ref{e3.5}) -- (\ref{e2.8})
are the necessary and sufficient conditions for the invariance of the
partition function under the local moves in the space ${\bf\tilde\Sigma}$.
In the next section, we shall provide the general solution of these
conditions.

\section{General Solutions}
Let $A$ be an associative semisimple $*$-algebra over the field of real
numbers $\R$, with the $*$-operation $\sigma: A \to A$, with  $\sigma^2 =id$
and $\sigma(ab)=\sigma(b)\sigma(a)$. Further, suppose that $A$ is equipped
with an inner product $ \br~,~\kt:A\times A\to \R $ and $\sigma$ is
self-adjoint with respect to this inner product.

In an arbitrary basis $\{ \phi_a~:~a=1,\cdots,N \}$, $\sigma$ is expressed by
a matrix ($\sigma_a^{~b})$, i.e: $\hat\phi_a=\sigma\phi_a=
\sigma_a^{~b}\phi_b $, and the conditions on $ \sigma $ take the following
form:
        \bea
        \sigma_a^{~b} \sigma_b^{~c}= \delta _a^c\,~~~~~~~~~~&&{\rm
        (involutiveness)}\;,
        \label{g11}\\
        C_{ba}^{~~c}\sigma_c^{c'}=
        \sigma_a^{a'}\sigma_b^{b'} C_{a',b'}^{c'}\,&&{\rm
        (antihomomorphism)}\;,
        \label{g12}\\
        \sigma_{ab}=\sigma_{ba}\,~~~~~~~~~~~&&\mbox{(self-adjointness)}\;.
        \label{g13}
        \eea
Note that
        $$\sigma_{ab} =\br\phi_a,\sigma \phi_b\kt=\br\phi_a,\sigma_{b}^{~b'}
        \phi_{b'}\kt=\sigma_{b}^{~b'}g_{ab'}=\sigma_{ba}\;, $$
also
        $$ \sigma^a_{~b} =
        g^{aa'}\sigma_{a'b}= g^{aa'}\sigma_{ba'} = \sigma_{b}^{~a}\;. $$
One can use Eqs.~(\ref{g11}) and (\ref{g13}) to write Eq.(\ref{g12}) in
the following equivalent form
        \be
        C_{ba}^{~~c} =
        \sigma_a^{~a'}\sigma_b^{~b'} \sigma_{c'}^{~c} C_{a',b'}^{~~c'}\;.
        \label{g14}
        \ee
Defining the metric as before, i.e., according to Eq.~(\ref{e2.8}),
we find that Eqs.~(\ref{g12}) and (\ref{g14}) are precisely the
necessary relations (\ref{e3.5}) and (\ref{e3.6}) for the formulation
of LTFT on arbitrary (not necessarily orientable) compact surfaces.
In fact, the relation with $*$-algebras can be seen quite naturally,
if one translates Fig.~\ref{f2.4} into the dual language (Fig.~\ref{f16}),
where a vertex shows the fusion of two elements of the algebra. In the
remainder of this article, we shall use a single line, rather than a
double line, to indicate a propagator and a single line with a dot to
indicate a twisted propagator, for simplicity.
        \begin{figure}
        \vspace{3in}
        \caption{Flipping transformation in ${\bf T_g}^*$.}
        \label{f16}
        \end{figure}

In view of these considerations, we have proven:
        \begin{itemize}
        \item[] {\bf Theorem~2}: {\em
        There is a one-to-one correspondence between the set of all LTFT's
        on two-dimensional compact surfaces (orientable or not) defined as
        above, and the set of all  semisimple real  associative $*$-algebras.}
        \end{itemize}
We conclude this section by recalling a couple of examples of associative
real $*$-algebras:
        \begin{itemize}
        \item[1)] Let $A$ be the algebra of real n-dimensional matrices
        $ M_n(\R) $ with the inner product $\br a,b\kt=tr(ab^t)$ and $\sigma$
        be the transpose operation $\sigma(a)=a^t$. A natural basis of
        $M_n(\R)$ is provided by the matrices $E_{ij}$ with $i,j=1,\cdots,n$
        defined by $(E_{ij})_{kl}:=\delta_{ik}\delta_{jl}$. We then have
        $\br E_{ij},E_{kl}\kt=g_{ij,kl}=N\delta_{ik}\delta_{jl}$,
        $g^{ij,kl}={1\over N}\delta^{ik} \delta^{jl}$, and $\sigma_{ij,kl}
        =N\delta_{il}\delta_{jk}$.
        \item[2)] Let $A=\R(G)$ be the group ring of a finite group $G$.
        For any two elements $a$ and $b$ of $G$, we define $\br a,b\kt=
        tr[\pi(a)\pi(b)]$ where $\pi$ denotes the regular representation of
        $G$, and induce an inner product on $\R(G)$ by linear extension.
        We also choose the $*$-operation to be the (linear extension of the)
        group inversion, $\sigma(a):=a^{-1}$. Then, it is easy to check that
        $\sigma$ is self-adjoint:
                \bea
                \br a,\sigma(b)\kt&=&tr[\pi(a)\pi(b^{-1})]\nn\\
                                  &=&[\pi(a)]^c_{~d}
                                  [\pi(b^{-1})]^d_{~c}\nn\\
                                  &=&C_{ad}^{~~c} C_{b^{-1}c}^{~~~~d}\nn\\
                                  &=&\delta({ad,c})\delta({b^{-1}c,d})\nn\\
                                  &=&\delta({ab^{-1}c,c})  \nn\\
                                  &=&\vert G \vert\delta_{a,b}
                                  \:=\: \br\sigma(a),b\kt\;,\nn
                \eea
        where $\delta(a,b):=\delta_{ab}$ is the kronecker delta function,
        i.e.,
                $$\delta(a,b):=\left\{\begin{array}{ccc}
                1&{\rm if}&a=b\\
                0&{\rm if}&a\neq b\,.
                \end{array}\right.$$
        \end{itemize}
\section{Physical Observables and Correlation Functions}
Let ${\bf \Sigma}$  be a (compact and connected) surface with an
$n$--component boundary. The boundary of ${\bf\Sigma}$ is homeomorphic to the
union of $n$ disjoint circles. Although ${\bf\Sigma}$ itself may not be
orientable, each component of its boundary may be oriented. Let us assign
the color indices $a_1,a_2,\cdots,a_n$ to the $n$ circles comprising the
boundary. We denote such a surface and a locally oriented triangulation
of it by ${\bf\Sigma}_{a_1,\cdots,a_n}$ and  ${\bf\Sigma}_{a_1,\cdots,a_n
}^{\alpha,k}$, respectively. We shall define the partition function,
${\cal Z}({\bf\Sigma}_{a_1,\cdots,a_n}^{\alpha,k})$, such that it will be
completely independent  of the triangulation and will depend only on
the color indices and the orientations of the boundary components. For
definition of the partition function we use exactly the same
set of rules as for the closed surfaces plus the following:
        \begin{itemize}
        \item[]{\em
Every boundary element with index $a$, whose orientation is (in)compatible
with that of the triangle adjacent to it, corresponds to a (twisted) untwisted
external line in the dual diagram (Fig.~\ref{f5.1}). Two different surfaces
are glued along their common boundary when the orientations of the boundaries
are opposite.}
        \end{itemize}
                \begin{figure}
                \vspace{1.5in}
                \caption{Triangle adjacent to a boundary component.}
                \label{f5.1})
                \end{figure}

We define the insertion of the operators $O_a$ ($a=1,2,\cdots,N$)
into the correlation functions as creating a loop boundary
with a fixed color index $a$ and summing over all other color indices of
the triangulation. We denote the correlation functions of $O_{a_1},\cdots,
O_{a_n}$ on a closed surface ${\bf\Sigma}$ by $\br O_{a_1}\cdots
O_{a_n}\kt_\Sigma$\,. Next, we prove:
        \begin{itemize}
        \item[]
        {\bf Theorem~3}: {\em
        The value of  ${\cal Z}({\bf\Sigma}_{a_1,\cdots,a_n}^{\alpha,k})$
        is independent of the triangulation, i.e.,
        ${\cal Z}={\cal Z}({\bf\Sigma}_{a_1,\cdots, a_n})$.}
        \item[]{\bf Proof}: We should only take care of the triangles
        adjacent to the external lines. Consider a flipping in the triangle
        adjacent to a boundary component (Fig.~\ref{f5.2}).
                \begin{figure}
                \vspace{1.5in}
                \caption{Flipping a triangle adjacent to a boundary component.}
                \label{f5.2}
                \end{figure}
        In the dual diagram this flipping is demonstrates also by
        Fig.~\ref{f16}. We know that due to Eqs.~(\ref{e3.5}) and (\ref{e3.6}),
        the partition function is invariant under such moves. In
Fig.~\ref{f5.2},
        we may also consider other possibilities for the orientations of the
        boundary components and the triangles, and see that the invariance of
        the correlation functions imposes no extra conditions besides
        Eqs.~(\ref{e3.5}) and (\ref{e3.6}).
        \end{itemize}

Note however that the correlation functions are invariant under
a reversal of the orientation of all the boundary components. This marks
a $\Z_2$--symmetry of our construction. In particular, this implies that
the one-point functions do not depend on the orientation of the boundary.
This is due to the fact that although one can compare two different
orientations of a given boundary component, one cannot compare the
orientations of two different boundary components. Thus, it is impossible
to assign an intrinsic value ($\pm$) to a given orienatation. This then means
that for a fixed set of indices on the $n$ boundary components of
${\bf\Sigma}$, one can define $2^{n-1}$ different correlation functions.
In the next section, we shall see how one can obtain all these $2^{n-1}$
different correlation functions from the knowledge of only one of them.

In the remainder of this section, we present some explicit calculations.

\subsection*{Calculation of Correlation Functions}
In the following we pursue the calculation, in general terms and without
specifying the underlying algebra, of the following quantities:
        \begin{itemize}
        \item[] The partition function of
                \begin{itemize}
                \item[A -] the sphere,
                \item[B -] the projective plane,
                \item[C -] the Klein bottle,
                \end{itemize}
        \item[] the one-point functions on
                \begin{itemize}
                \item[D -] the sphere,
                \item[E -] the projective plane,
                \item[F -] the Klein bottle,
                \end{itemize}
        \item[] and finally,
                \begin{itemize}
                \item[G -] the two-point function on the sphere,
                \item[H -] the three-point function on the sphere, and
                \item[I -] the partition and correlation functions on
                arbitrary compact surfaces.
                \end{itemize}
        \end{itemize}
We shall see that observables D, E, F, and H can be used as building
blocks for calculation of all correlation functions on arbitrary compact
surfaces, i.e., I. In our graphical calculations, we shall use the identities
depicted in Fig.~\ref{f21}.
        \begin{figure}
        \vspace{4in}
        \caption{Graphical identities}
        \label{f21}
        \end{figure}

Next, we pursue the computation of:
\vspace{-.1cm}
\subsection*{A- Partition function of the sphere $S^2$}
We can always normalize the partition function of the sphere to unity. For
future use we present in Fig.~\ref{f20},
        \begin{figure}
        \vspace{3in}
        \caption{A triangulation of the sphere and its dual diagram.}
        \label{f20}
        \end{figure}
the simplest triangulation of the 2-sphere together with its dual
graph.\footnote{Note that the multiple arrows on the edges of some of
triangles are used to mean that they are to be identified. They are not to be
confused with the single arrows which specify the orientations of the
boundary components.}
By performing second Matveev move in the dual graph, we see that the dual
diagram of $S^2 $ is a circle. Therefore we have:
        $${\cal Z}(S^2)={\cal Z}(\bigcirc\hspace{-2.5mm}|\hspace{2.5mm} )=
        {\cal Z}(\bigcirc)=1\;.$$
\subsection*{B-Partition function of the projective plane $\R P^2$}
A simple triangulation of the projective plane and the corresponding dual
graph is shown in Fig.~\ref{f22}.
        \begin{figure}
        \vspace{2in}
        \caption{A triangulation of $\R P^2$ and its dual graph.}
        \label{f22}
        \end{figure}
In order to compute the partion function, first we simplify the dual diagram
by performing the first and then the second Matveev moves in the lower area.
The result is demonstrated in Fig.~\ref{f23}.
        \begin{figure}
        \vspace{2in}
        \caption{A simplified dual diagram for $\R P^2$.}
        \label{f23}
        \end{figure}
{}From the latter diagram we obtain:
        \be
        {\cal Z}(\R P^2)=C_{ca}^{~~b}C_{db}^{~~a}\sigma^{cd}\;.
        \label{h.14}
        \ee

\subsection*{C- Partition function of the Klein bottle ${\cal K}$}
Fig.~\ref{f24} shows a triangulation of the Klein bottle and its dual
diagram, where we have also indicated how to simplify the dual diagram using
Matveev moves.
        \begin{figure}
        \vspace{2in}
        \caption{A triangulation of the Klein bottle and its dual diagram.}
        \label{f24}
        \end{figure}
In view of Fig.~\ref{f24}, we obtain :
        \be
        {\cal Z}({\cal K})=C_{b'c'}^{~~a} C_{cba}\sigma^{cc'}\sigma^{bb'}
        \label{h.15}
        \ee

\subsection*{D- One-point function on the sphere (disk)}
Removing the interior of a circle from the sphere and fixing an index $a$
on the circle (Fig.~\ref{f25}),
        \begin{figure}
        \vspace{1.5in}
        \caption{A triangulation of the disk.}
        \label{f25}
        \end{figure}
we obtain the one-point function on the sphere , which is topologically a
disk. Hence, we have
        \be
        \br O_a\kt_{S^2} =C_{ab}^{~~b}\;.
        \label{h16}
        \ee

\subsection*{E- One-point function on the projective plane (Mobius strip)}
The simplest triangulation of the one-point function on the projective plane
is shown in Fig.~\ref{f26}. This is obtained by removing the interior of a
circle from  $\R P^2$. Topologically, this corresponds to the Mobius strip.
In view of Fig.~\ref{f26},
        \begin{figure}
        \vspace{1.5in}
        \caption{A triangulation of the Mobius strip.}
        \label{f26}
        \end{figure}
we have:
         \be
         \br O_{a}\kt_{\R P^2}=C_{abc}\,\sigma^{bc}\;.
         \label{h17}
         \ee

\subsection*{F- One-point function on the Klein bottle}
In order to compute the one-point function on the Klein bottle, we cut a disk
in Fig.~\ref{f24}, and obtain Fig.~\ref{f27}.
        \begin{figure}
        \vspace{2in}
        \caption{A triangulation of the one-point function on Klein bottle.}
        \label{f27}
        \end{figure}
The latter leads to:
        \be
        \br O_{a}\kt_{{\cal K}}=C_{ab}^{~~c} C_{cd'}^{~~e'}C_{ed}^{~~b}
        \sigma^{dd '} \sigma^{~e}_{e'}\;.
        \label{h18}
        \ee
\subsection*{G-  Two-point functions on the sphere}
According to the orientations of the boundaries there are two different
two-point functions on the sphere, depicted in Fig.~\ref{f28} which we call
$\eta_{ab}$ and $\xi_{ab}$.
        \begin{figure}
        \vspace{3in}
        \caption{A triangulation of the two-point functions on $S^2$.}
        \label{f28}
        \end{figure}
One can find the simplest triangulation of $\eta_{ab} $ and $\xi_{ab}$
by representing both of them as rectangles with two idendified sides.
According to Fig.~\ref{f28}:
        \bea
        \eta_{ab}&=&C_{ac}^{~~d} C_{db}^{~~c}\;,
        \label{h19}\\
        \xi_{ab}&=&\eta_{a}^{~b'}\sigma_{b'b}\;.
        \label{h20}
        \eea
Gluing two $\eta$'s or two $\xi$'s, one can verify the following identities:
        \be
        \eta_a^{~b}\eta_b^{~c}=\eta_a^{~c}\;,~~~~\eta_a^{~b}\xi_b^{~c}=
        \xi_a^{~b}\eta_b^{~c}=\xi_a^{~c}\;,~~~~\xi_a^{~b}\xi_b^{~c}=
        \eta_a^{~c}\;.
        \label{h21}
        \ee
In fact, the first identity is the same as in the orientable case. The
remaining two identities are consequences of Eq.~(\ref{h20}). The
significance of Eqs.~(\ref{h21}) will be emphasized below.

In Ref.~[FHK], it is shown that $\eta$ is a projection onto the center $Z(A)$
of the algebra $A$, i.e., $\eta_a^{~b}C_{bcd}=\eta_a^{~b}C_{bdc}$, which
implies:
        \bea
        \forall\phi\in A~:&&\eta\phi\in Z(A)\;,\nn\\
        \forall\tilde\phi\in Z(A)~:&&\eta\tilde\phi=\tilde\phi\;.
        \label{h22}
        \eea
Moreover, in view of Eq.~(\ref{h20}), $\xi$ also acts as a projector to
the center $Z(A)$, although it is not a proper projection due to
the last relation in (\ref{h21}).

Note that by gluing $\xi_{ab}$ to any boundary component of the surface,
we can change its prescribed orientation. Thus the correlation functions
corresponding to different assignments of the orientation to the boundary
components may be obtained in this way from a given one.

At this stage, we would like to relabel the indices of the basis
$\{\phi_a~:~a=1,\cdots,N\}$ of $A$ in such a way that the first $M$
indices label the basis of $Z(A)$:
        \be
        A=\bigoplus_{a=1}^N\R\phi_a=Z(A)\oplus Z^c(A):=
        \left(\bigoplus_{\alpha=1}^M
        \R\phi_\alpha\right)\oplus\left(\bigoplus_{i=M+1}^N\R\phi_i\right)\;.
        \label{h23}
        \ee
Since $\eta=(\eta_a^{~b})$ is a projector onto $Z(A)$ and Eq.~(\ref{h22})
holds, $\eta$ takes the following form in the new basis:
        \bea
        (\eta_{ab})&=&\left[
        \begin{array}{cc}
        \eta_{\alpha\beta}=g_{\alpha\beta}&0\\
        0&0\end{array}\right]\;,\nn\\
        (\eta_a^{~b})&=&\left[
        \begin{array}{cc}
        \eta_\alpha^{~\beta}=\delta_\alpha^{~\beta}&0\\
        0&0\end{array}\right]\;,
        \label{h24}\\
        (\eta^{ab})&=&\left[
        \begin{array}{cc}
        \eta^{\alpha\beta}=g^{\alpha\beta}&0\\
        0&0\end{array}\right]\;.\nn
        \eea

An interesting observation is that $\sigma$ induces a $\Z_2$--grading
of the center $Z(A)$, althought it does not induce such a grading on
the whole algebra $A$. Thus, we have:
        \be
        Z(A)=Z^+(A)\oplus Z^-(A)=\left(\bigoplus_{\alpha^+=1}^{M_1} \R
        \phi_{\alpha^+}
        \right)\oplus\left(\bigoplus_{\alpha^-=M_1+1}^M \R\phi_{\alpha^-}
        \right)\;,
        \label{h26}
        \ee
where $\sigma\phi_{\alpha^\pm}=\pm\phi_{\alpha^\pm}$, and
        \bea
        Z^+(A)~Z^+(A)&\subset&Z^+(A)\nn\\
        Z^+(A)~Z^-(A)&\subset&Z^-(A)
        \label{x1}\\
        Z^-(A)~Z^-(A)&\subset&Z^+(A)\;.\nn
        \eea

\subsection*{H - Three-point functions on the sphere}

The simplest triangulation for the three-point function on sphere,
with the prescribed orientations as shown in Fig.~\ref{f31},
      \begin{figure}
      \vspace{3in}
      \caption{A triangulation of a three-point function on $S^2$ with
      a prescribed orientation on the boundary components.}
      \label{f31}
      \end{figure}
leads to a dual diagram consisting of three $\eta$'s joint at a vertex [FHK].
Thus, we have:
        \be
        N_{abc}:=\br O_aO_bO_c\kt= \eta_a^{~a'}\eta_b^{~b'}\eta_c^{~c'}
        C_{a'b'c'}\;.
        \label{h29}
        \ee
Note that in view of Eqs.~(\ref{h24}),
        \be
        N_{\alpha\beta\gamma}=C_{\alpha\beta\gamma}\;.
        \label{h27}
        \ee
Other choices of orientations on the boundary components correspond to
replacing some of $\eta$'s by $\xi$'s in Eq.~(\ref{h29}).

Since every insertion of operator $O_a$ (to obtain a multi-point function) is
necessarily subject to the projection by $\eta$ or $\xi$, the following
theorem [FHK] also generalizes to the case considered in this paper.
        \begin{itemize}
        \item[]{\bf Theorem~4}: {\em The set of physical observables is in
        one-to-one correspondence with the center $Z(A)$ of the
        the real associative $*$-algebra $A$ associated with the LTFT.
        In particular, the number of the independent physical operators
        is equal to the dimension of $Z(A)$.}
        \end{itemize}

In view of Eq.~(\ref{h27}) and the $\Z_2$--grading of $Z(A)$ demonstrated
by Eqs.~(\ref{x1}), we can regard $O_{\alpha^+}$ and $O_{\alpha^-}$ as
``bosonic'' and ``fermionic'' observables. This terminology is motivated
by the follwing ``selection rules'':
        $$ N_{\alpha^+\beta^+\gamma^-}=N_{\alpha^-\beta^-\gamma^-}=0\;.$$

\subsection*{I - Case of general compact surfaces}
To compute the correlation functions of other compact surfaces, we
appeal to the following result:
        \begin{itemize}
        \item[] {\bf Theorem~5}: {\em
        The one-point functions on the sphere $D_\alpha$, the Klein
        bottle ${\cal K}_\alpha$, the projective plane ${\cal M}_\alpha$,
        and the three-point function on sphere $N_{\alpha\beta\gamma}$
        can be used as building blocks to find any correlation function
        on any compact connected surface by gluing.}
        \item[] {\bf Proof}:
        First note that by gluing a disk $D_\alpha$ to a three-point
        function $N_{\alpha\beta\gamma}$ on the sphere, one obtains the
        two-point function $\eta_{\alpha\beta}$ on the sphere. Gluing
        $\eta_{\alpha\beta}$ to $N_{\alpha\beta\gamma}$, one obtains a
        handle operator which is used in the construction of surfaces
        of higher genus. Furthermore, gluing $N_{\alpha
        \beta\gamma}$ to any $n$--point function yields an $(n+1)$--point
        function on the same surface.
        Next, one can glue ${\cal M}_\alpha$ (resp.\ ${\cal K}_\alpha$)
        to the $(n+1)$--point function on a genus $g$ orientable surface
        $\Sigma_g$ to obtain the $n$--point function on the
        non-orientable surface $\Sigma_g\mbox{\#}\R P^2$
        (resp.~ $\Sigma_g\mbox{\#}{\cal K}$).
        According to the classification theorem for two-dimensional
        surfaces [M], this exhausts all the possibilities
        of the multi-point functions on arbitrary compact surfaces.
        \end{itemize}

These considerations can be expressed in an algebraic language by defining
the matrices:
         $$ (N_\beta)_\alpha^{~\gamma}:=N_{\alpha\beta}^{~~\gamma}\;,$$
the vectors $\omega$, ${\cal M}$, and ${\cal K}$ with components:
        $$ \omega_\alpha:=tr(N_\alpha)\;,~~~~~{\cal M}_\alpha\;,~~~~~
        {\cal K}_\alpha\;,$$
respectively, and the matrix:
        $$\tilde{N}:=\sum_{\alpha=1}^M\omega_\alpha N_\alpha\;.$$
Denoting by $g$ the genus of the surface, we will then have
for the orientable surfaces $\Sigma_{g}$:
        \bea
        \br O_{\alpha_1}\cdots O_{\alpha_n}\kt_{g=0}&=&\left(
        N_{\alpha_2}N_{\alpha_3}\cdots N_{\alpha_{n-1}}\right)_{\alpha_1}^{
        \alpha_n}\;,
        \label{h30}\\
        \br O_{\alpha_1}\cdots O_{\alpha_n}\kt_{g=1}&=&tr~\left(
        N_{\alpha_1}N_{\alpha_2}\cdots N_{\alpha_n}\right)\;,
        \label{h31}\\
        \br O_{\alpha_0}\kt_g&=&( N_{\alpha_0}N_{\alpha_1}\cdots
        N_{\alpha_g})\omega_{\alpha_1}\omega_{\alpha_2}
        \cdots\omega_{\alpha_g}\;,\nn\\
        &=&\left( \tilde{N}^{g-1}\omega\right)_{\!\alpha_0}\;,
        \label{h32}\\
        \br O_{\alpha_1}\cdots O_{\alpha_n}\kt_g&=&
        \br O_{\alpha_1}\cdots O_{\alpha_n}O_{\alpha_{n+1}}\kt_{g=0}
        \br O_{\alpha_{n+1}}\kt_g\;,\nn\\
        &=& \left( N_{\alpha_2}\cdots N_{\alpha_n} \tilde{N}^{g-1}\omega
        \right)_{\!\alpha_1}\;,
        \label{h33}\\
        {\cal Z}(\Sigma_g)&=&\omega_{\alpha_1}\cdots\omega_{\alpha_g}
        \br O_{\alpha_1}\cdots O_{\alpha_g}\kt_{g=0}\:=\:
        \omega^t\, \tilde{N}^{g-2}\omega\;,
        \label{h34}
        \eea
and for non-orientable surfaces:
        \bea
        {\cal Z}(\Sigma_g \#{\cal K})&=&{\cal K}_\alpha\br O_\alpha\kt_g
        \:=\: {\cal K}^t \tilde{N}^{g-1}\omega\;,
        \label{h35}\\
        {\cal Z}(\Sigma_g \#\R P^2)&=& {\cal M}_\alpha\br O_\alpha\kt_g
        \: =\: {\cal M}^t\, \tilde{N}^{g-1}\omega\;,
        \label{h36}\\
        \br O_\alpha\kt_{\Sigma_g\#{\cal K}}&=&
        ({\cal K}^t\tilde{N}^g)_\alpha\;,
        \label{h37}\\
        \br O_\alpha\kt_{\Sigma_g \#\R P^2}&=&
        ({\cal M}^t\tilde{N}^g)_\alpha\;,
        \label{h38}\\
        \br O_{\alpha_1}\cdots O_{\alpha_n}\kt_{\Sigma_g\#{\cal K}}&=&
        \br O_{\alpha_1}\cdots O_{\alpha_{n+1}}\kt_{g=0}
        \br O_{\alpha_{n+1}}\kt_{\Sigma_g\#{\cal K}}\;,\nn\\
        &=&\left( N_{\alpha_2}\cdots N_{\alpha_n}\tilde{N}^g{\cal K}\right)_{
        \alpha_1}\;,
        \label{h39}\\
        \br O_{\alpha_1}\cdots O_{\alpha_n}\kt_{\Sigma_g \#\R P^2}&=&
        \left( N_{\alpha_2}\cdots N_{\alpha_n}\tilde{N}^g{\cal M}\right)_{
        \alpha_1}\;,
        \label{h40}
        \eea
where the superscript ``$t$'' stands for the ``transpose''.

\section{Example: The Group Ring $A=\R(G)$}
In this section we deal with the special case where
$A=\R[G]:=\bigoplus_{a\in G}\R a$, is a group ring associated with a finite
group $G$ of order $|G|$. In this case, one has:
        \be
        C_{ab}^{~~c}=\delta(ab,c)\;.
        \label{e4.1}
        \ee

The group ring $A$ is naturally a real $*$-algebra with the $*$-operation
given by linear extention of:
        \be
        \sigma(a):=a^{-1}\;,~~~~~\forall a\in G\;.
        \label{e4.2}
        \ee
Using Eqs.~(\ref{e4.1}) and~(\ref{e4.2}), we have:
        \bea
        g_{ab}&=&|G|\delta(a,b^{-1})\;,
        \label{e4.3}\\
        C_{abc}&=&|G|\delta(abc,1)\;,
        \label{e4.4}\\
        \sigma_{ab}&=&|G|\delta(a,b)\;.
        \label{e4.5}
        \eea
Similarly,  we find
        \bea
        g^{ab}&=&\frac{1}{|G|}\delta(a,b^{-1})\;,
        \label{e4.6}\\
        \sigma^{ab}&=&\frac{1}{|G|}\delta(a,b)\;,
        \label{e4.7}\\
        \sigma^{a}_{~b}&=&\sigma^{~a}_b\:=\:
        \delta(a,b^{-1})\;,
        \label{e4.8}
        \eea
In view of these equations, we may easily compute:
        \be
        \eta_{ab}=\br O_a O_b\kt_0=\frac{|G|}{h_{[a]}}
        \delta([a],[b^{-1}])\;.
        \label{e4.9}
        \ee
Here, $[a]$ denotes the conjugacy class of $a$, i.e.,
        \[
        [a]:=\{b\in G~:~ b=g\,a\,g^{-1}\;,~g\in G\}\;,\]
and $h_{[a]}$ is the number of elements of $[a]$. Furthermore, we have
        \bea
        \eta^{~b}_{a}&=&\eta_{ac}g^{cb}=\frac{1}{h_{[a]}}\delta(
        [a],[b])\;,
        \label{e4.10}\\
        \xi_{ab}&=&\eta^{~c}_{a}\sigma_{cb}\:=\:
        \frac{|G|}{h_{[a]}}\delta([a],[b])\;,
        \label{e4.11}\\
        \xi^{~b}_a&=& \xi_{ac}g^{cb}\:=\:
        \frac{1}{h_{[a]}}\delta([a],[b^{-1}])\;.
        \label{e4.12}
        \eea

Next, we consider some specific examples:
        \begin{itemize}
        \item[1.] The partition function for the sphere $S^2$:
                \be
                {\cal Z}(S^2)={\cal Z}(\bigcirc\hspace{-2.5mm}|\hspace{2.5mm})
                =C_{abc}C_{a'b'c'}g^{aa'}g^{bb'}g^{cc'}=
                \frac{1}{|G|}\sum_{a,b,c}
                \delta(abc,1)\delta(a^{-1}c^{-1}b^{-1},1)=1\;.
                \label{i50}
                \ee
        \item[2.] One-point function on $S^2$ (The disk ($D$)):
                \be
                \br O_a\kt_{S^2}= C_{ab}^{~~b}=\sum_b\delta(ab,b)
                =|G|\delta(a,1)\;.
                \label{e4.14}
                \ee
        \item[3.] The partition function for the projective plane $\R P^2$:
                \bea
                {\cal Z}(\R P^2)&=& C_{ab}^{~~c}C_{dc}^{~~b}\sigma^{da}\nn \\
                &=&\frac{1}{|G|}\sum_{a,b,c,d}
                \delta(ab,c)\delta(dc,b)\delta(d,a)\nn\\
                &=&\frac{1}{|G|}\sum_{a}
                \delta(a^2,1)\;.\nn
                \eea
        The sum in the latter equation can be split into a sum over the
        distinct conjugacy classes $[b]$, followed by a sum over the
        elements belonging to each class, $a\in [b]$. Then, in view of
        the identity:
                $$ \sum_{a\in [b]}\delta(a^2,1)=
                \frac{|G|}{h_{[b^2]}}
                \delta([b^2],1)\;,$$
        one finally has:
                \be
                {\cal Z}(\R P^2)=\sum_{[b]}
                \frac{1}{h_{[b^2]}}\delta([b^2],1)\;.
                \label{e4.15}
                \ee
        \item[4.] One-point function on $\R P^2$ (the Mobius strip
        (${\cal M}$)):
                \bea
                \br O_a\kt_{\R P^2}&=&{\cal M}_a\:=\:
                C_{abc}\sigma^{cb}=C_{ab}^{~~c}\sigma_c^{~b}\nn\\&=&
                \sum_{b,c}
                \delta(ab,c)\delta(c,b^{-1})\nn\\&=&\sum_b\delta(ab,b^{-1})=:
                G_a^{1/2}\;.
                \label{e4.16}
                \eea
        Here, $G_a^{1/2}$ is the number of elements of $G$ whose square
        equals $a$. Note that $G_a^{1/2}$ is a function of $[a]$. To see this
        suppose that $b_i$,~$i=1,\cdots,G_a^{1/2}$ are such that $b_i^2=a$.
        Then for all $g\in G,~~ b_i':=g\,b_ig^{-1}$ have the property that
        $b_i^{'2}=g\,a\,g^{-1}=a'\in [a]$. Thus,
$G_{gag^{-1}}^{1/2}=G_a^{1/2}$.
        \item[5.] The partition function of the Klein bottle (${\cal K}$):
                \be
                {\cal Z}({\cal K})= C_{b'c'}^{~~a}C_{cba}\sigma^{cc'}
                \sigma^{bb'}=\sum_{[a]}
                \frac{1}{h_{[a]}}\delta([a],[a^{-1}])\;.
                \label{e4.17}
                \ee
        \item[6.] One-point function on ${\cal K}$:
                \be
                \br O_a\kt_{{\cal K}}=
C_{ab}^{~~m}C_{md'}^{~~e'}C_{ed}^{~~b}\sigma^{dd'}\sigma^e_{~e'}
                =\sum_{[b]}\frac{1}{h_{[b]}}\delta([ab],[b^{-1}])\;.
                \label{e4.18}
                \ee
        \end{itemize}

We conclude this section emphasizing the fact that all the
correlation functions are functions of the conjugacy classes. This is to
be expected since the physical observables are related to the center of
the algebra and the center is spanned by the conjugacy classes. Furthermore,
the physical observables being functions only of the conjugacy classes
can be expressed in terms of the characters of the irreducible representations
of the group.




\section{Conclusion}
In this article, it is shown how in two dimensions one can formulate state
sums on non-orientable compact manifolds. Pursuing the same approach as
in the treatment of the orientable case, one encounters the problem of
the lack of a canonical orientation for the non-orientable surfaces. This
manifests itself in the lack of a canonical prescription for the assignment
of ordered $C_{abc}$'s to the triangles of a given triangulation. The solution
offered above involves the following three steps:
        \begin{itemize}
        \item[1)]
        Introduction of locally oriented triangulations,
        \item[2)]
        Generalization of the Matveev moves, i.e., inclusion of flipping
        transformation.
        \item[3)]
        Employing the $*$-structure of real associative $*$-algebras
        to ensure the topological invariance of the partition and correlation
        functions.
        \end{itemize}
Thus, at a more fundamental level, the $\Z_2$--obstruction of
non-orientability leads to the requirement of the existence of a
$*$-structure for the underlying algebra of any LTFT on non-orientable
manifolds.

A similar problem exists in three dimensions where adjacent tetrahedra
with incompatible orientations are present in any triagulation.
It seems that our approach may be applied to this case, as well.

\section*{Acknowledgments}
V.~K.\ wishes to thank Amir.~M.~Ghezelbash for fruitful discussions.
We would also like to thank Aref Mostafazadeh for his help in drawing
figures and Kamran Kaviani and Aziz Shafikhani for their assistance
with the computer editing.

\newpage

\section*{References}
{\small
\begin{itemize}
\item[{\bf [A]}~~] Atiyah, M.~: ``Topological quantum Field theories,''
Publ.~Math.\
I.~H.~E.~S. {\bf 68}, 175 (1989).
\item[{\bf [B]}~~] Bachas, C.~ and Petropouls, M.~: Commun.~Math.~Phys. {\bf
152}, 191
(1993).
\item[{\bf [D]}~~] Durhuus, B.~: ``A discrete approach to topological quantum
field
theories,'' J.~Geom.\ and Phys.\ {\bf 11}, 155 (1993).
\item[{\bf [DJN]}] Durhuus, B., Jakobsen, H.~ and Nest, R.~: ``Topological
quantum
field theories from generalized $6j$--symbols, Rev.~Math.~Phys.
{\bf 5}, 1 (1993).
\item[{\bf [DW]}~] Dijkgraaf, R.~and Witten, E.~: ``Topological gauge theories
and
group cohomology,'' Preprint,~IASSN-HEP-89/33.
\item[{\bf [FHK]}] Fukuma, M., Hosono, S.~and Kawai, H.~: ``Lattice topological
field theory in two dimensions,'' Commun.~Math.~Phys. {\bf 161}, 157 (1994).
\item[{\bf [GSW]}] Green, M., Schwartz, J., Witten, E.~: ``Superstring
Theory,''
Cambridge University Press, Cambridge (1987).
\item[{\bf [KS]}~] Karowski, M.~and Schrader, R.~: ``A combinatorial approach
to
topological quantum field theories and invariants of graphs,''
Commun.~Math.~Phys. {\bf 151}, 355 (1993).
\item[{\bf [M]}~~] Massey, W.~: ``Algebraic Topology,'' Springer-Verlag, Berlin
(1990).
\item[{\bf [MS]}~] Moore, G.~and Seiberg, N.~: ``Lectures presented at Trieste
Spring
School (1989).
\item[{\bf [TV]}~] Turaev, V.~and Viro, O.~: ``State sum invariants of
3-manifolds
and quantum $6j$--symbols,'' Topology {\bf 31}, 865 (1992).
\item[{\bf [W1]}~] Witten E.~: ``Topological quantum field theory,''
Commun.~Math.~Phys. {\bf117}, 353 (1988).
\item[{\bf [W2]}~] Witten E.~: ``Quantum Field Theory and the Jones
Polynomial,''
Commun.~Math.~Phys. {\bf 121}, 351 (1989).
\end{itemize}}

\end{document}